# Analysis of the HiSCORE Simulated Events in TAIGA Experiment Using Convolutional Neural Networks

**Anna Vlaskina**[a] **and Alexander Kryukov**[a,b]

[a]*M.V. Lomonosov Moscow State University, Leninskie Gory, Moscow, 119991, Russian Federation*

[b]*Department, University,*
*Street number, City, Country Skobeltsyn Institute of Nuclear Physics, 1(2) Leninskie gory, Moscow 119991, Russian Federation*

*E-mail:* vlaskina.aa18@physics.msu.ru, kryukov@theory.sinp.msu.ru

TAIGA is a hybrid observatory for gamma-ray astronomy at high energies in range from 10 TeV to several EeV. It consists of instruments such as TAIGA-IACT, TAIGA-HiSCORE, and others. TAIGA-HiSCORE, in particular, is an array of wide-angle timing Cherenkov light stations. TAIGA-HiSCORE data enable to reconstruct air shower characteristics, such as air shower energy, arrival direction, and axis coordinates. In this report, we propose to consider the use of convolution neural networks in task of air shower characteristics determination. We use Convolutional Neural Networks (CNN) to analyze HiSCORE events, treating them like images. For this, the times and amplitudes of events recorded at HiSCORE stations are used. The work discusses a simple convolutional neural network and its training. In addition, we present some preliminary results on the determination of the parameters of air showers such as the direction and position of the shower axis and the energy of the primary particle and compare them with the results obtained by the traditional method.

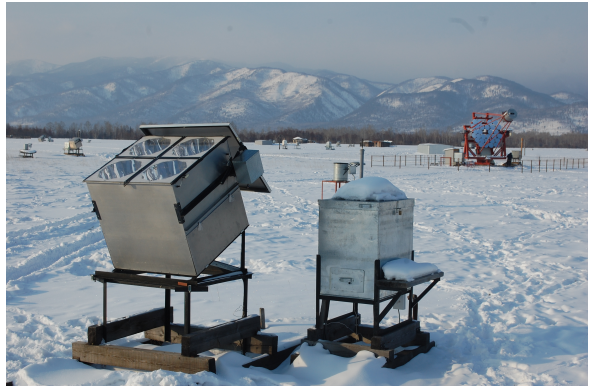

**Figure 1:** HiSCORE station.

## 1. Introduction

Gamma astronomy and cosmic-ray physics are rapidly developing areas of physics, one of the goals of which is to search for the origins of high-energy gamma rays and cosmic rays to answer some fundamental questions about the Universe. There are many different instruments for studying gamma rays and cosmic rays, such as IceCube [1], the neutrino observatory, or VERITAS [2], a Cherenkov telescope array. Combining several approaches to gamma-ray and cosmic-ray physics could produce impressive results.

At high energies, cosmic rays can be analyzed by studying the air showers they generate in the atmosphere. When the original particle reaches the atmosphere, it produces a massive cascade of secondary particles that emit Cherenkov light. Such a cascade of particles is called an Extensive Air Shower (EAS).

**TAIGA experiment**

TAIGA (Tunka Advanced Instrument for cosmic rays and Gamma Astronomy) [3] is a hybrid observatory for cosmic ray physics. It is a complex system for ground based astronomy from a few TeV to several PeV. The observatory consists of IACTs, wide-angle Cherenkov stations, scintillation arrays, electron and muon detectors. Use of several different types of detectors enable multimessenger analysis of EAS.

TAIGA-HiSCORE is an array of wide-angle Cherenkov stations. Stations are located at a distance of 75 m and 150 m from each other. Due to the high sensitivity of the stations and the ability to measure the signal registration time at different stations, the accuracy of determining the EAS parameters is higher than that of many other instruments. [4]

## 2. Methods of EAS parameters determination

A classical approach to EAS parameter reconstruction is based on Hillas parameters analysis [5]. Hillas parameters are characteristics of events registered by Cherenkov stations. The event has the form close to ellipse, and Hillas parameters are ellipse characteristics, such as its size, orientation, gravity center, length, and width. This method allows not only air shower parameters reconstruction but discrimination of gamma and hadron events as well.

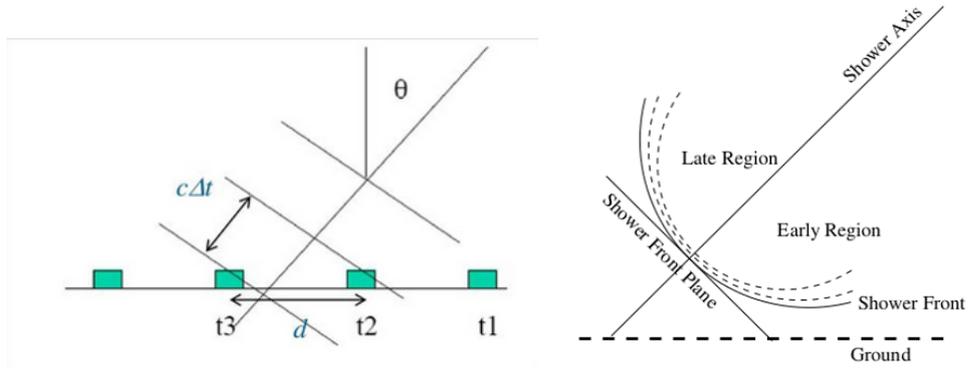

**Figure 2:** The picture on the left schematically shows how to determine an azimuth angle. Knowing the period of time between signal detection of each station and distance between the stations, azimuth angle $\theta$ can be found from a triangle. The picture on the right is a schematic representation of the shower front.

HiSCORE provides information about the exact time of the signal registration. Having this information enables us to reconstruct the EAS front in terms of the signal registration time. As shown in Figure 2, the EAS front reaches two different stations for a known period of time. Knowing signal registration times and coordinates of the stations, we can derive azimuth angle $\theta$. As for a polar angle $\phi$, it can be determined knowing air shower front plane location. However, when applying these direction determination methods, we assume that the shower front is flat. The need to consider the EAS front flat could be a serious limitation of these methods.

In proposed method, we represent arrays of data obtained from several HiSCORE stations as images. The image can consist of signal timing data or from data on the signal amplitudes. While working with times of signal registration, we are looking for a time isoline and use it to restore the direction of the EAS axis. For time isoline determination, deep learning technologies can be used. In particular, we propose using neural networks for EAS parameter reconstruction.

Artificial Neural Network (ANN) – a mathematical model capable of learning by examples. Such a structure can solve problems of classification or regression. During training, our model receives EAS data that can be obtained during an actual experiment. At the output of the neural network, we expect to receive the EAS parameters. Training is done by the backpropagation method. [**?**] The error computed in the output layer is sent back to the network to refine the output for each neuron, which is fed back to the neuron in the output layer for more accurate output than before. Thus, the local minimum of the error function is sought.

A convolutional neural network is a special kind of neural network for processing data with a grid topology [9], such as images. Convolutional networks are neural networks that use convolution instead of the general operation of matrix multiplication in at least one layer. Due to this sparseness, the convolutional neural network requires less memory, using only meaningful features. The convolutional neural network also extracts information about the distance between objects, this information is used for more accurate pattern recognition.

The work [6] has shown that ANN is a useful tool for gamma astronomy. The model, which is a combination of a convolutional neural network and a recurrent neural network, performed the classification task on IACT (Imaging Atmospheric Cherenkov Telescope) events and showed promising results. Moreover, in work [7], the neural network solves not only the problem of classifying particles but also the problem of determining the EAS energy, which is a linear regression problem. We suggest that other EAS parameters can be determined with high accuracy using neural networks. In this report, we propose using convolutional neural network for determining polar and azimuth angles of air shower axis. To solve this problem, we use data on the times of signal registration at HiSCORE stations, presented in the form of images.

## 3. Methodology and results

**Data**

To train our model, we use Monte-Carlo simulated events. The dataset was prepared using CORSIKA software [10]. Monte-Carlo simulation made possible modeling events with set EAS parameters, such as air shower axis coordinates, primary particle type, polar and azimuth angels of air shower axis, and primary particle energy. In this work, only proton events were used in a dataset. Each event contains of data on the signal amplitudes and signal registration times by HiSCORE stations. On early stages of our work we only use the information about signal timing. Arrays of this data can be represented as images, allowing the effective use of a convolutional neural network for EAS parameters determination. Using signal registration times, the polar and azimuth angles of the EAS axis are determined.

The dataset contained data from 44 Cherenkov stations. Figure 3 shows how HiSCORE stations are located. In each event, the signal is detected by at least 4 stations. Since the neural network accepts as input a rectangular image, additional points with zero values have been added to the array. It must be taken into account, that the real locations of Cherenkov stations differ from the regular grid within a few meters. Therefore,interpolation of time values for each station was performed. Trilinear interpolation for each grid point was performed using the values at the nearest three points. The point for which we calculate the refined value of the function is located inside a triangle of three other points. It is enough to perform linear interpolation twice: first for the intermediate point, then for the desired point. As a result, corrections for time values were taken into account, as they can play a significant role in model training. In addition to that, a simple data normalization by minimum value was implemented for each event.

11400 images for training set and 816 for testing set were selected from the original dataset. With the help of oversampling, we increased our training set to 45600. The oversampling was carried out by rotating the original images by 90 degrees.

**Neural network architecture**

The scheme of our CNN model is presented in Figure 4. Our model is a sequential convolutional neural network with four convolutional layers and five dense layers on top of it. In the convolutional layers feature extraction takes place, while in the dense layers linear regression is performed. We chose ReLU (fig.5) as an activation function on all layers since the values of angles are positive. Table 1 details the CNN parameters. To implement the model, the work uses the TensorFlow [11] and Keras [12] libraries, used for machine learning on Python 3.1.1. This model could be adjusted to determine other parameters, such as energy and air shower axis coordinates.

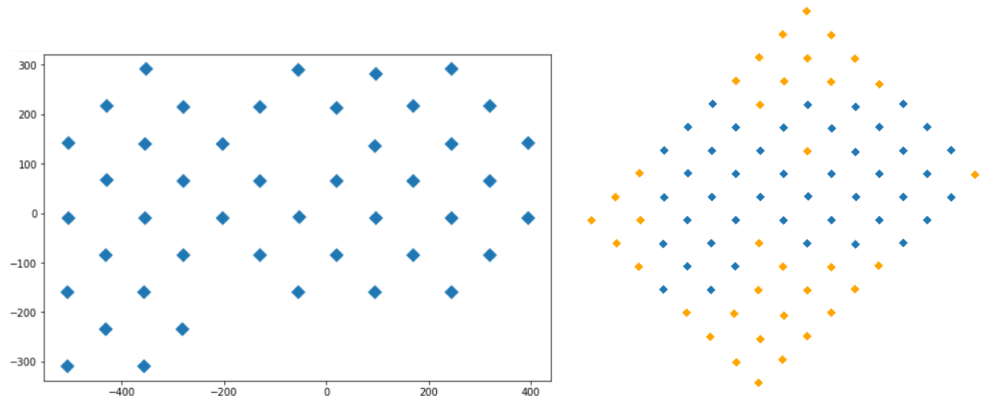

**Figure 3:** HiSCORE 2D coordinates. The mesh can be converted to rectangular from triangular if we rotate coordinate axes 45°. We add zero-value points to the array to make it rectangular. Orange dots on the right image represent added points.

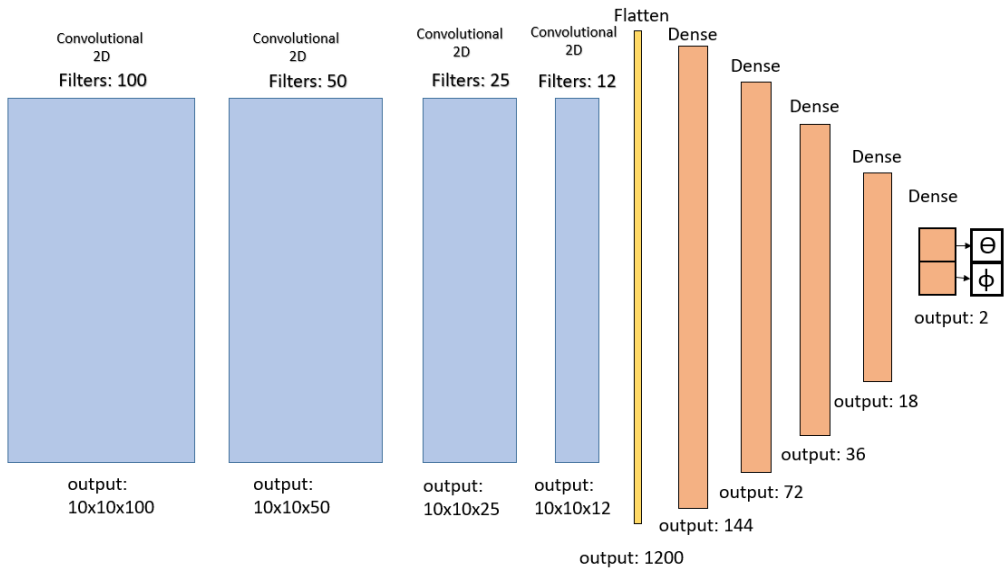

**Figure 4:** The proposed neural network architecture.

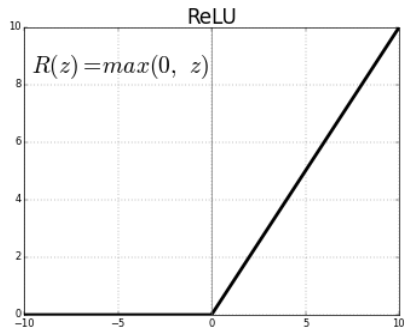

**Figure 5:** ReLU activation function.

**Table 1: Model architecture and training parameters**

| | |
|---|---|
| CNN kernel size | 3x3 |
| Activation function | ReLU |
| Learning rate | 0.001 |
| Batch size | 10 |
| Epoch number | 70 |
| Total number of parameters: | 246,402 |
| Optimizer | ADAM (Adaptive moment estimation) |
| Loss function | Mean Sqared Error |

**Performance evaluation**

To estimate the error during and after training, we used the MSE function:

$$MSE = \frac{1}{n} \sum_{i}^{n} (y_{true} - y_{pred})^2$$

Overall, the model demonstrated total loss of 2.80 for two parameters. Let's take a closer look at results for each parameter. We will estimate the deviations from the expected values using the mean absolute error function (MAE):

$$MAE = |y_{true} - y_{pred}|$$

| Parameter | Mean absolute error |
|---|---|
| azimuth angle $\theta$ | 0.9° |
| Polar angle $\phi$ | 1.2° |

Errors for events with different numbers of triggered stations can vary significantly. Let's divide our dataset into events with the number of triggered stations more than ten and less than ten. Mean absolute errors for events with number of triggered stations below ten and above ten are shown in the table below:

| | Parameter | Mean absolute error |
|---|---|---|
| **Below 10:** | azimuth angle $\theta$ | 1.1° |
| | Polar angle $\phi$ | 1.5° |
| | Parameter | Mean absolute error |
| **Above 10:** | azimuth angle $\theta$ | 0.8° |
| | Polar angle $\phi$ | 1.1° |

## 4. Conclusion and future work

A trained convolutional neural network predicts polar and azimuth angles with an accuracy of 1.5°. These results are promising, however, our aim is to reduce the error to 0.3° − 0.4°, what has been achieved by conventional methods. The deep learning approach has proven to be flexible and fast in execution, so further investigations are needed. To further our research we are planning to use more complex network architectures, such as GoogLeNet [13] or ResNet [14]. We intend to use both timing arrays and arrays of signal amplitudes recorded by the stations simultaneously. Future work will explore deep learning applications to the EAS energy and the position of the EAS axis determination as well.

## Acknowledgments

The authors are grateful to the TAIGA collaboration for useful discussions and the provided model data.